\documentclass[
  aps,
  prb,
  twoside,
  twocolumn,
  showpacs,
  floatfix,
  superscriptaddress,
  10pt,
  preprintnumbers,
  citeautoscript,
]{revtex4-1}

\usepackage{amsmath}
\usepackage{comment}
\usepackage{glossaries}
\usepackage{graphicx}
\usepackage{here}
\usepackage[utf8]{inputenc}
\usepackage{times}
\usepackage{units}

\newacronym{cbm}{CBM}{conduction band minimum}
\newacronym{dft}{DFT}{density functional theory}
\newacronym{dos}{DOS}{density of states}
\newacronym{dwf}{DWF}{Debye-Waller factor}
\newacronym{ks}{KS}{Kohn Sham}
\newacronym{hr}{HR}{Huang-Rhys}
\newacronym{hrf}{HRF}{Huang-Rhys factor}
\newacronym{ifc}{IFC}{interatomic force constant}
\newacronym{ipr}{IPR}{inverse participation ratio}
\newacronym{pl}{PL}{photoluminescence}
\newacronym{psb}{PSB}{phonon sideband}
\newacronym{puc}{PUC}{primitive unit cell}
\newacronym{si}{SI}{supplemental information}
\newacronym{vbm}{VBM}{valence band maximum}
\newacronym{zpl}{ZPL}{zero phonon line}
\newacronym{spe}{SPE}{single photon emitter}

\newcommand{\vsineg}{V$^{-1}_{\text{Si}}$}
\newcommand{\vsidblneg}{V$^{-2}_{\text{Si}}$}
\newcommand{\vsi}{V$_\text{Si}$}
\newcommand{\vc}{V$_\text{C}$}
\newcommand{\vccsi}{V$_{\text{C}}$C$_{\text{Si}}$}
\newcommand{\vccsipos}{{V$_{\text{C}}$C$_{\text{Si}}$}$^{+1}$}
\newcommand{\vcvsi}{V$_{\text{C}}$V$_{\text{Si}}$}
\newcommand{\vcvsineut}{{V$_{\text{C}}$V$_{\text{Si}}$}$^{0}$}
\newcommand{\ncvsi}{N$_{\text{C}}$V$_{\text{Si}}$}
\newcommand{\ncvsineg}{{V$_{\text{Si}}$N$_{\text{C}}$}$^{-1}$}

\newcommand{\qunit}{\AA$^2$amu}

\newcommand{\vb}{\mathbf{v}}

\newcommand{\inner}[2]{\langle#1|#2\rangle}

\usepackage[colorlinks=true]{hyperref}
\hypersetup{
    unicode=false,          % non-Latin characters in Acrobat’s bookmarks
    pdftoolbar=true,        % show Acrobat’s toolbar?
    pdfmenubar=true,        % show Acrobat’s menu?
    pdffitwindow=false,     % window fit to page when opened
    pdfstartview={FitH},    % fits the width of the page to the window
    pdfnewwindow=true,      % links in new PDF window
    colorlinks=true,        % false: boxed links; true: colored links
    linkcolor=blue,         % color of internal links (change box color with linkbordercolor)
    citecolor=blue,         % color of links to bibliography
    filecolor=blue,         % color of file links
    urlcolor=blue           % color of external links
}

\newcommand{\specf}{$S(\omega)$}

\usepackage{xcolor}
\usepackage{booktabs}
\AtBeginDocument{
\heavyrulewidth=.08em
\lightrulewidth=.05em
\cmidrulewidth=.03em
\belowrulesep=.65ex
\belowbottomsep=0pt
\aboverulesep=.4ex
\abovetopsep=0pt
\cmidrulesep=\doublerulesep
\cmidrulekern=.5em
\defaultaddspace=.5em
}

\begin{document}
\begin{abstract}
Silicon carbide with optically and magnetically active point defects offers unique opportunities for quantum technology applications.
Since interaction with these defects commonly happens through optical excitation and de-excitation, a complete understanding of their light-matter interaction in general and optical signatures in particular is crucial.
Here, we employ quantum mechanical density functional theory calculations to investigate the photoluminescence lineshapes of selected, experimentally observed color centers (including single vacancies, double vacancies and vacancy-impurity pairs) in 4H-SiC.
The analysis of zero-phonon lines as well as Huang-Rhys and Debye-Waller factors is accompanied by a detailed study of the underlying lattice vibrations.
We show that the defect lineshapes are governed by strong coupling to bulk phonons at lower energies and localized vibrational modes at higher energies.
Generally, good agreement to the available experimental data is obtained, and thus we expect our theoretical work to be beneficial for the identification of defect signatures in the photoluminescence spectra and thereby advance the research in quantum photonics and quantum information processing.
\end{abstract}
\title{
Photoluminescence Lineshapes for Color Centers \texorpdfstring{\\}{}
in Silicon Carbide from Density Functional Theory Calculations}

\newcommand{\aaltophysics}{Department of Applied Physics, Aalto University, P.O. Box 11100, 00076 Aalto, Finland}
\newcommand{\chalmersphys}{Department of Physics, Chalmers University of Technology, Gothenburg, Sweden}
\newcommand{\aaltocentre}{QTF Centre of Excellence, Department of Applied Physics, Aalto University, FI-00076 Aalto, Finland}
\newcommand{\helmholtz}{Helmholtz-Zentrum Dresden-Rossendorf, Institute of Ion Beam Physics and Materials Research, 01328 Dresden, Germany}
\newcommand{\oulu}{Microelectronics Research Unit, University of Oulu, 90014 Oulu, Finland}

\author{Arsalan Hashemi}
\affiliation{\aaltophysics}
\author{Christopher Linder\"alv}
\affiliation{\chalmersphys}
\author{Arkady V. Krasheninnikov}
\affiliation{\aaltophysics}
\affiliation{\helmholtz}
\author{Tapio Ala-Nissila}
\affiliation{\aaltocentre}
\author{Paul Erhart}
\affiliation{\chalmersphys}
\author{Hannu-Pekka Komsa}
\affiliation{\aaltophysics}
\affiliation{\oulu}

\maketitle
\section{Introduction}
During the last decade, improvements in the fabrication techniques of silicon carbide (SiC) a material with a wide electronic band gap of \unit[3.2]{eV} \cite{Fisher1990} have made it possible to produce high-quality samples \cite{Weber2010, Melinon2007} with control over types and concentrations of color centers in this system, making SiC attractive for applications in nano-photonics, electronics, and spintronics \cite{Castelletto_2020,Lohrmann_2017,Aharonovich2016}.
These centers can potentially be used as single photon sources and spin-photon interfaces \cite{Lukin2020, Awschalom2018, Son2020, Eisaman2011, O'Brien2009, Lodahl2015, Gao2015}, and thus could play a central role in future quantum technologies \cite{Atature2018, Hesen2015, Bogdanov17, Borreti2014}.
Nowadays, it is also possible to prepare SiC with largely isolated color centers which allows investigations of the properties of specific types of color centers \cite{Lei2017, Christle2017}.

SiC exists in several polytypes, with 3C, 4H, and 6H being the most common ones.
The majority of the research on color centers for single photon emitters and spin qubits reported in the literature has been carried out on 4H-SiC, as it is easy to synthesize in high quality and with a low density of stacking faults.
4H-SiC, which has a high Debye temperature of \unit[1200]{K} \cite{Lou2009} and strong second-order optical nonlinearity \cite{Wu2008}, consists of four SiC sheets stacked in ABCB order along the $c$-axis.
It contains two nonequivalent hexagonal ($h$) and quasi-cubic ($k$) sites for each atom.
In addition, different charge states can be favorable for some color centers \cite{Defo_2018, Gordon2015prb}.
As a result, 4H-SiC can host several defects with distinct \gls{pl} peaks within the band gap.
The most important ones are:
(i) The negatively charged silicon vacancy (\vsineg) in $h$ or $k$ sites with \gls{zpl} in the range of \unit[1.352--1.445]{eV} \cite{Shang2020, Udvarhelyi2020, Widmann2014, Sorman2000}.
It offers two excitation possibilities at each vacancy site.
(ii) The neutral carbon-silicon divacancy (\vcvsi) can occur in four distinct configurations ($hh$, $hk$, $kh$, and $kk$)
and exhibits \gls{zpl} in the range of \unit[1.095--1.15]{eV} \cite{Davidsson_2018, Kohel_2011, Flak2014, Christle2017, Wolfowicz2017}.
Both \vsi{} and \vcvsi{} are paramagnetic color centers and exhibit long spin coherence times \cite{Christle2015, Udvarhelyi2020}. 
(iii) The carbon antisite-vacancy pair (\vccsi{}) 
has been characterized in $-1$, $0$, and $+1$ charge states\cite{Umeda2006, Umeda2007, Kriszti2015, Castelletto2013, Steeds2009}.
(iv) The extrinsic negatively charged silicon vacancy-nitrogen pair (\ncvsineg) is equivalent to the famous NV center in diamond \cite{Csoren2017,Hong2013mrs}
and emits light in the \unit[1]{eV} range \cite{Wang_prb2020,BardelebenPRB2015, Zargaleh2018, Zargaleh2018prb}.

The experimentally measured \gls{pl} spectra are defect specific and originate from the combination of electronic transitions with phonons.
While purely electronic transitions give rise to the so-called \glspl{zpl}, transitions that involve phonon creation or annihilation lead to the so-called \gls{psb}, which is located at lower (higher) energies than the \gls{zpl} in photon emission (absorption) spectra.
From the \gls{zpl} and \gls{psb}, one can extract the \gls{hrf} and the \gls{dwf} \cite{Walker1979},
which are important parameters to assess the suitability of defects for quantum applications.
The \gls{dwf} describes the relative ratio of light emitted in the \gls{zpl} relative to the total emitted light and the \gls{hrf} describes the average number of phonons involved in the emission.
Thus, a larger \gls{dwf} and a smaller \gls{hrf} are obtained when the electron-phonon interaction is weaker and indicate a lower information leakage to the environment, which is important for the usage of color centers in quantum information science and nanosensing \cite{Bradac2019, Maze2008, Johnson2015}.

Although all these defects have been the subject of several publications, the experimental results are usually focused on an individual color center.
The only well-reported signatures of defects are the spectrally narrow emission spectra at a certain wavelength at low temperatures \cite{Shang2020, Udvarhelyi2020, Widmann2014, Sorman2000, Davidsson_2018, Kohel_2011, Flak2014, Christle2017, Wolfowicz2017, Wang_prb2020,BardelebenPRB2015, Zargaleh2018, Zargaleh2018prb, Umeda2006, Umeda2007, Kriszti2015, Castelletto2013, Steeds2009}.
Electronic structure calculations can be extremely useful to obtain detailed insight and understanding of defects, at a level inaccessible to experiment.
Accordingly, they have been routinely used to analyze, e.g., charge transition levels, spin states, and \glspl{zpl} \cite{Davidsson_2018, Yan2020, Defo_2018, Kobayashi2019, Gordon2015prb, Torpo2002}.
It was, however, only very recently \cite{Shang2020, Udvarhelyi2020} that the \glspl{psb} of \vsi{} were calculated and compared to experimental data.
While \glspl{dwf} were reported, the contributions from different vibrational modes to them remain poorly understood.

Here, to ameliorate this situation, we provide a comprehensive and systematic analysis of the vibrational signatures of all of the aforementioned defects.
Using \gls{dft} in combination with the generating function approach, we obtain the \glspl{psb}, \glspl{zpl}, \glspl{dwf}, and \glspl{hrf} and determine spin states and charge localization for each defect.
These results are compared to experimental results, where available.
We then analyze the vibrational modes via the electron-phonon coupling spectral function and separate localized vibrational modes from bulk phonons.
Finally, we discuss our results in the context of technological applications.

\section{Methodology}

\subsection{Generating function approach}

\Gls{pl} lineshapes, and the \gls{psb} in particular, are modelled via the generating function approach as described in Refs.~\onlinecite{Lax52, Miyakawa1970, Alkauskas2014}.
To model the \gls{psb}, the electron-phonon interaction has to be accounted for.
The interaction can be formalized using the partial \gls{hrf} defined as 
\begin{equation}
    S_\lambda = \omega_\lambda Q_\lambda^2 \bigg/ 2\hbar,
\end{equation}
where $\hbar$ is the reduced Planck constant and the configurational coordinate $Q_\lambda$ for an optical process is defined as
\begin{equation}
 Q_{\lambda} = \sum_{\alpha} \sqrt{m_{\alpha}} \langle(\bf{R}_{e, \alpha} - {\bf{R}}_{g, \alpha}) | {\bf{u}}_{\alpha, \lambda}\rangle. 
\end{equation}
Here, ${\bf{R}}_{g}$ and ${\bf{R}}_{e}$ are ground and excited state atomic coordinates,
while ${\bf{u}}_{\alpha, \lambda}$ indicates the normalized displacement vector
corresponding to mode $\lambda$ with frequency $\omega_{\lambda}$ and $m_{\alpha}$ is mass of atom $\alpha$.
The total \gls{hrf} and $Q^{2}$ are defined as $S = \sum S_\lambda$ and $ Q^{2} = \sum Q^{2}_\lambda$, respectively.
They provide measures for the average number of phonons involved in the emission and the difference between the initial and final state geometries.
From the \gls{hrf}, the \gls{dwf} can be defined as $\text{DWF}=\exp(-S)$ \cite{Alkauskas2014}.
The spectral function that underlies the computation of the \gls{psb} is
\begin{equation}\label{eq:s}
S(\omega) = \sum_{\lambda}S_\lambda\delta(\omega - \omega_{\lambda}).  
\end{equation}
Once the electron-phonon spectral function is computed,
the spectral distribution function can be determined as
\begin{equation}
    A(\omega) = \int dt \exp(S-S(t))\exp(i\omega t),
\end{equation}
where $S(t)$ is the Fourier transform of $S(\omega)$. The emission intensity is proportional to $\omega^3 A(\omega)$. 

\subsection{Computational details}

Electronic structure calculations were performed within the framework of \gls{dft} using the projector augmented wave method \cite{Blo94, KreJou99} as implemented in the Vienna ab-initio simulation package \cite{KreFur96, KreHaf93}.
For structural relaxations and phonon calculations, the PBEsol exchange correlation functional was used \cite{PBEsol}.
To correct for the well-known band gap error intrinsic to semi-local exchange-correlation functionals such as PBEsol, additional calculations were carried out using the HSE06 hybrid exchange correlation functional \cite{HSE06}.
To obtain accurate transition energies and \glspl{zpl}, we (i) used PBEsol-optimized structures, (ii) scaled the lattice parameters to be consistent with the values from HSE06, and (iii) performed total energy calculations using HSE06 without further relaxation.
The plane-wave energy cutoff was set to \unit[400]{eV}.
Ionic optimization was performed until forces were
smaller than \unit[10]{meV/\AA} and the break condition for
the electronic self-consistent loop was set to \unit[$10^{-6}$]{eV}.
The first Brillouin zone of the primitive unit cell was sampled using a $10\times 10\times4$ $\boldsymbol{k}$-point grid.

To model defects, $5\times 5\times 2$ supercells containing up to 400 atoms were used.
The Brillouin zone was sampled using a zone centered $2 \times 2 \times 2$ grid and the $\Gamma$-point in PBEsol and HSE06 calculations, respectively.
These supercells correspond to a defect concentration of about $2.5 \times 10^{20}\,\text{cm}^{-3}$.
Spin-polarized calculations were performed for all charge states.
To model optical transitions, we constrained the partial occupancies of the \gls{ks} levels, promoting one electron from the highest occupied state to the next or second-next higher-energy state in the same spin channel.
If the electron was promoted to a doubly degenerate state, we fixed the occupancy weight for each state to one half to achieve faster convergence of the calculation \cite{Alkauskas2014}.

\begin{figure}%[ht!]
    \centering
 \includegraphics[width=\linewidth]{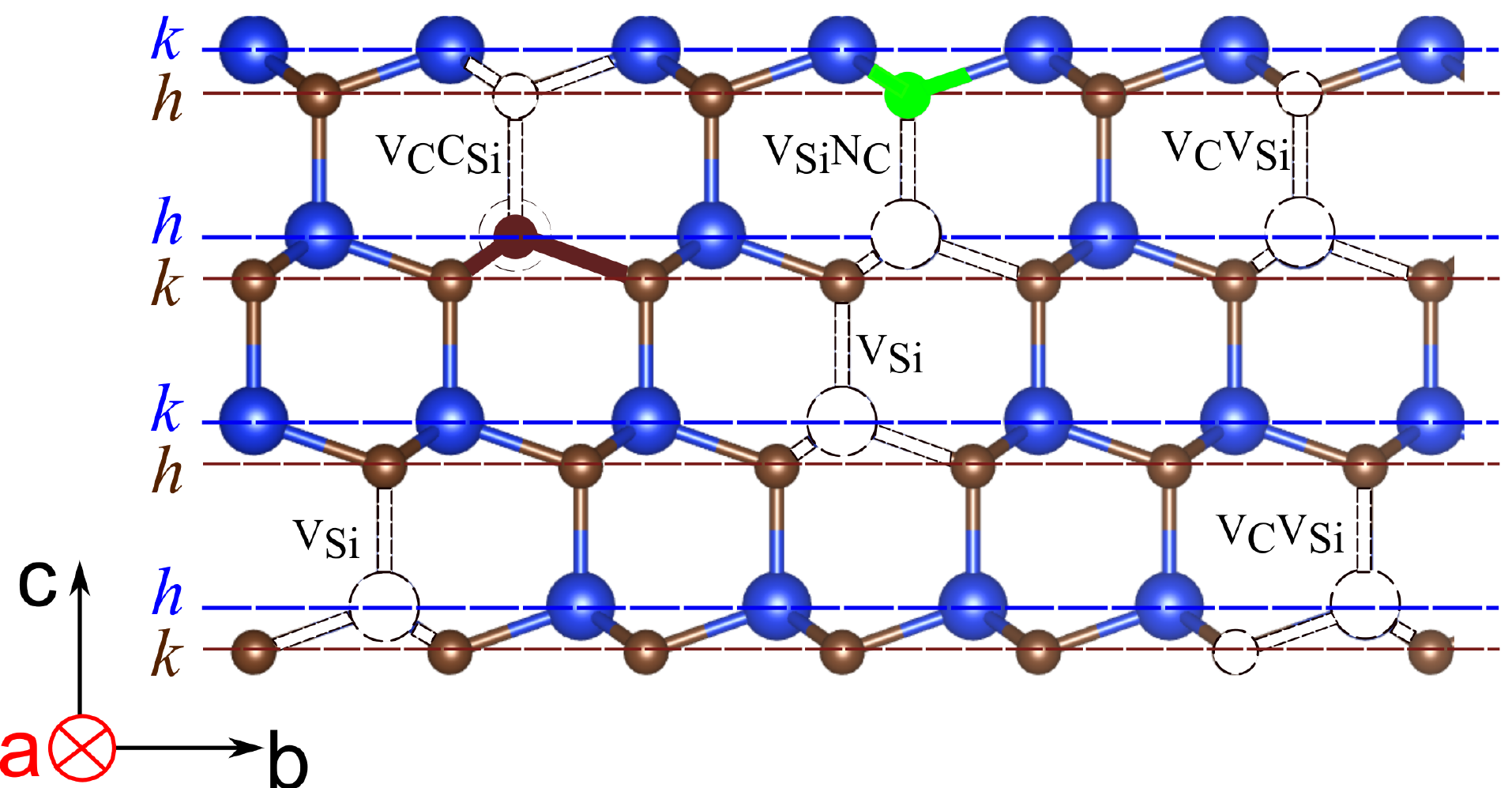}
    \caption{
        Schematic representation of defect configurations in 4H-SiC.
        Blue, brown, and green spheres denote silicon, carbon, and nitrogen atoms, respectively.
        Vacancy sites are shown as hollow spheres.
        The planes  labelled $h$ and $k$ indicate the symmetry of lattice sites.
    }
    \label{fig:def4hsic}
\end{figure}

\begin{figure*}[htbp]
    \centering
    \includegraphics[width=\linewidth]{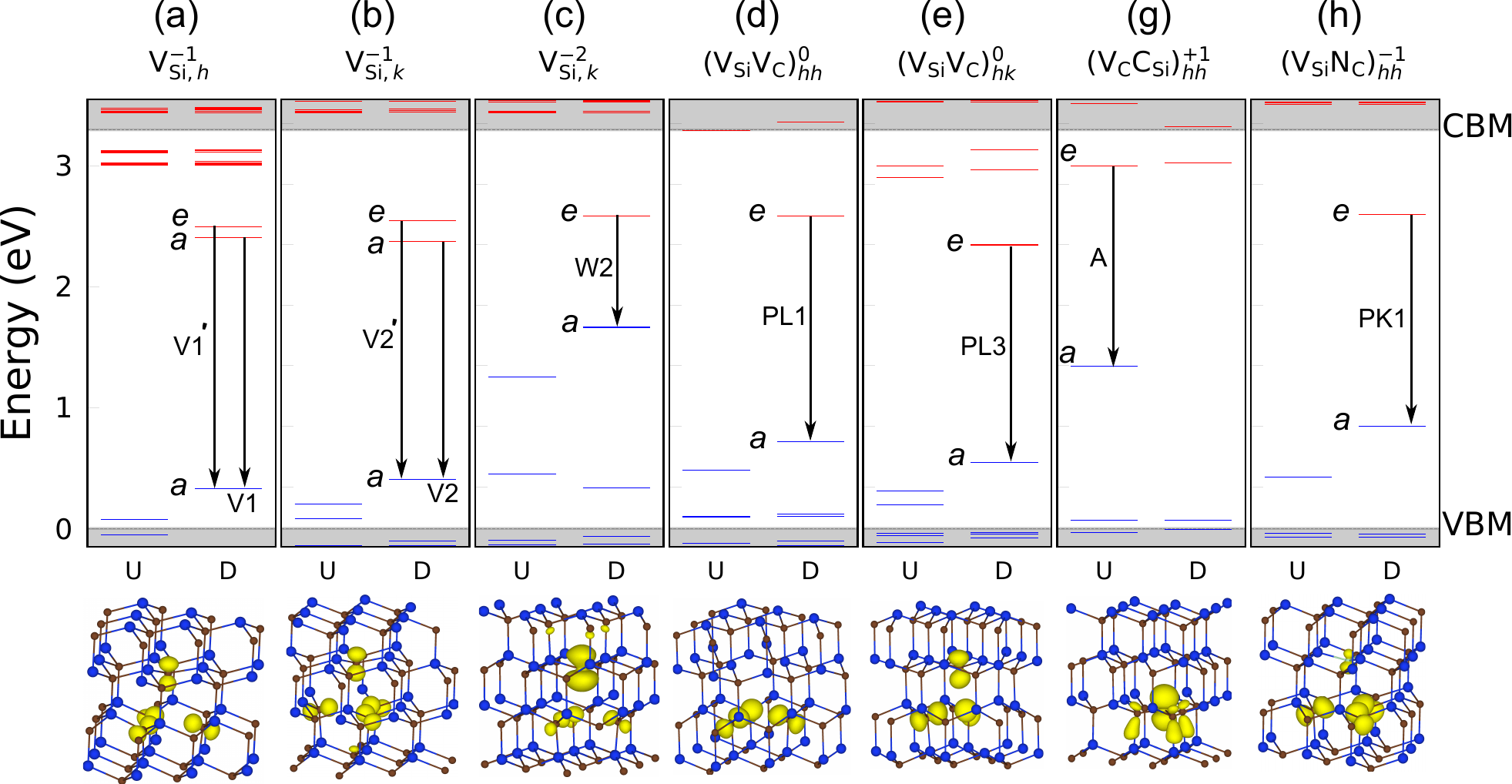}
    \caption{
        Top: Arrows show the optical excitation between the highest occupied (blue lines) and unoccupied \gls{ks} levels (red lines).
        Doubly degenerate and non-degenerate states are denoted by $e$ and $a$, respectively.
        After excitation, the electron decays to the ground electronic state by photon and phonon emission.
        The majority (spin-up) and minority (spin-down) channels are denoted by U and D, respectively.
        Bottom: The partial charge densities of the highest occupied state in the transition channel for the ground state configurations.
        Si, C, and N atoms are denoted by blue, brown and green balls.
    }
    \label{fig:transitions}
\end{figure*}

The defects considered in this work are illustrated in \autoref{fig:def4hsic}.
%????
For Si vacancies, we include the singly negative charge state (\vsineg{}) positioned at both h and k sites and doubly negative charge state (\vsidblneg{}) at k site \cite{Kobayashi2019, Gordon2015prb}.
%and positions at both h and k sites.
Regarding divacancies (\vcvsi{}), we study configurations comprising a Si vacancy on an $h$ site in combination with C vacancies on either $h$ or $k$ sites.
Finally, we include two variants of the NV-center in diamond, namely the positively charged \vccsi{} and the negatively charged \ncvsi{} complexes.
In total, there are thus seven centers and nine optical (charge neutral) transitions possible (\autoref{fig:transitions}).

Phonon calculations for large defect systems are computationally demanding.
To accelerate these calculations we used regression as implemented in the \textsc{hiphive} package \cite{HIPHIVE}.
For each transition, training data was prepared by calculating the forces for the ground state structure using up to 125 configurations with random displacements, which were drawn from a normal distribution with a standard deviation of  \unit[0.01]{\AA}.
The \glspl{ifc} were then reconstructed using a cluster basis set with second-order cutoffs of 4.2 to \unit[5.0]{\AA} and optimized by least-squares regression.
For all cases, the root-mean-square error over a holdout set that comprised 5\% of the total available data was \unit[17]{meV/\AA} or lower.
Finally, the \glspl{ifc} were used to compute phonon frequencies and eigenvectors with the \textsc{phonopy} package \cite{PHONOPY}.

\section{Results}
\label{sec:results}

\subsection{Electronic structures and excitations}

The 4H-SiC structure has hexagonal symmetry with the lattice constants $a$ and $c$.  
Our results obtained using the PBEsol functional for the lattice constants are $a=$\unit[3.08]{\AA} and $c=$ \unit[10.08]{\AA}.
The electronic band gap from HSE06 calculations is found to be $E_g=\unit[3.25]{eV}$.
These values are in good agreement with experimental data of $a=\unit[3.09]{\AA}$, $c=\unit[10.48]{\AA}$, and $E_g= \unit[3.2]{eV}$ \cite{Lebedev1999}.

The defects considered here exhibit at least one occupied and one unoccupied defect state in the band gap, comprising a two-level optical system that involves states localized in the proximity of the defect. In \autoref{fig:transitions} the electronic transitions are shown.
Transitions are labeled in accordance with previous experimental studies except for \vsidblneg{}, for which no established label exists.
All transitions were obtained by promoting an electron in the spin minority channels except for configuration A, for which the optical transition occurs in the majority spin channel.
The ground state electronic configurations exhibit spin states $S = 3/2$, 1, 1, $1/2$, and 1 for \vsineg{}, \vsidblneg{}, \vcvsineut{}, \vccsipos{}, and \ncvsineg{} centers, respectively.
These values were experimentally and theoretically confirmed \cite{Janzen2009, Kriszti2015, gali2010, Defo_2018, Wang_prb2020}.

In the electronic ground state of \vsineg{} (\autoref{fig:transitions}a-b), four C dangling bonds point to the vacant site with their localized orbitals.
\vsineg{} on an $h$ site is associated with transitions V1 and V1', which correspond to HOMO$\leftarrow$LUMO and HOMO$\leftarrow$LUMO+1 transitions, respectively.
Analogously, \vsineg{} on an $k$ site supports transitions V2 and V2'.
The two transitions are separated by \unit[116]{meV} and \unit[172]{meV}, respectively, at $h$ and $k$ sites.
The LUMO+1 states are doubly degenerate.
Here the excited state is constructed by equally occupying each degenerate orbital. 

Upon addition of one extra electron to \vsineg{} to obtain \vsidblneg{}, the electrons are still localized next to the vacant site (\autoref{fig:transitions}(c)).
The charge distribution is mostly localized on the C atom placed at the same crystallographic site as vacant Si.
In contrast to \vsineg{}, \vsidblneg{} can only host a single optical transition, labelled W2 for $k$ site.
Previous formation energy calculations showed that the doubly negative Si vacancy is stable in charge state $-2$ \cite{Defo_2018}, but the predicted \gls{pl} line has not yet been experimentally observed.
The W2 transition exhibit a \gls{zpl} energy of \unit[0.72]{eV}, which is in the telecom wavelength ($<$\unit[1]{eV}).

The neutral divacancies (\autoref{fig:transitions}d-e) can exhibit different symmetries depending on the local symmetry of the vacancy C and Si atoms.
In order to restrict ourselves, we consider the HOMO$\leftarrow$LUMO transitions on the divacancy in $hh$ and $hk$.
Based on the \gls{ks} spectra, additional transitions are possible in both $hh$ and $hk$ symmetry.
The HOMO$\leftarrow$LUMO transition on $hh$ is labelled PL1 and the HOMO$\leftarrow$LUMO transition on $hk$ is labelled PL3.
The electronic charge is equally distributed among the C atoms nearest to the silicon vacancy.
It is noteworthy that the three nearest Si and C neighbors to the C and Si vacancy, respectively, relax inward toward the respective vacant site by \unit[0.02]{\AA}.

The singly charged carbon vacancy-carbon antisite defect \vccsi$^{+1}$ supports one transition labelled A, which occurs between two localized levels in the majority spin channel (\autoref{fig:transitions}g).
Charge is mostly localized on the antisite carbon and to a lesser degree on its three nearest C neighbors.
With \unit[1.55]{\AA} the C--C bond length is about \unit[0.34]{\AA} shorter than Si--C bonds in the ideal lattice.

Finally, the negatively charged silicon vacancy adjacent to a nitrogen that substitutes carbon, \ncvsineg{}, which contains one additional electron compared to \vsineg{}, allows for a single optical transition based on the \gls{ks} spectrum (\autoref{fig:transitions}h), which is labelled PK1.
Only a very small part of the density of the unpaired electrons is found at the N atom.
The defect resembles the axial \vcvsi{} divacancy with the Si--N bond length being about \unit[0.11]{\AA} shorter than Si--C bonds in ideal lattice.

Our calculated \gls{zpl} energies differ by less than \unit[0.2]{eV} from the experimentally measured ones with the exception of the PK1 transition on \ncvsineg{} (\autoref{tab:info}), a level of accuracy that matches earlier calculations \cite{Davidsson_2018}.
Generally, the good agreement with experiment of the calculated \gls{zpl} energies, as well as some of the other defect properties to be considered in the following, validates the defect assignments given in the literature.

\begin{table}
\centering
\caption{
    Key properties from calculation and experiment for the defect transitions considered here.
}
\vspace{0.2cm}
%\resizebox{\columnwidth}{!}{
\label{tab:info}
\begin{tabular}{llccccccc}
    \toprule
    &
    & \multicolumn{2}{c}{\gls{zpl} (eV)}
    & \multicolumn{1}{l}{$Q^{2}$ (\AA$^{2}$\,amu)}
    & \multicolumn{1}{c}{\gls{hrf}}
    & \multicolumn{2}{c}{\gls{dwf} (\%)} \\
    & & HSE06 & Expt.                              & PBEsol & PBEsol & PBEsol & Expt. \\
    \midrule
    \vsineg{}
    & V1    & 1.57  & 1.44\cite{Shang2020}         & 0.62   & 2.78   &  6.17  & 6--8\cite{Shang2020} \\
    & V1' & 1.60  & 1.44\cite{Sorman2000}        & 0.66   & 2.77   &  6.23  & ---   \\
    & V2    & 1.24  & 1.35\cite{Shang2020}         & 0.64   & 2.82   &  5.94  & 9.0\cite{Udvarhelyi2020} \\
    & V2' & 1.28  & ---                                & 0.47   & 1.94   & 14.34  & --- \\[3pt]
    \vsidblneg{}
    & W2    & 0.72  & ---                                & 0.30   & 1.47   & 22.97  & --- \\[3pt]
    \vcvsi{}
    & PL1   & 1.14  & 1.10\cite{Kohel_2011}        & 0.71   & 2.75   &  6.39  & 5.3\cite{Christle2015} \\% $\pm$ 1.1\cite{Christle2015} \\
    & PL3   & 1.25  & 1.12\cite{Falk2013}          & 0.99   & 3.08   &  4.57  &  5.0\cite{Crook2020} \\[3pt]
    \vccsi{}
    & A     & 1.69  & 1.88\cite{Castelletto2013}   & 0.73   & 3.77   &  2.29  & --- \\[3pt]
    \ncvsi{}
    & PK1   & 1.28  & 1.00\cite{BardelebenPRB2015} & 0.70   & 2.64   &  7.09  & --- \\
    \bottomrule
\end{tabular}
%}
\end{table}

\subsection{Luminescence lineshapes}
\begin{figure*}
    \centering
    \includegraphics{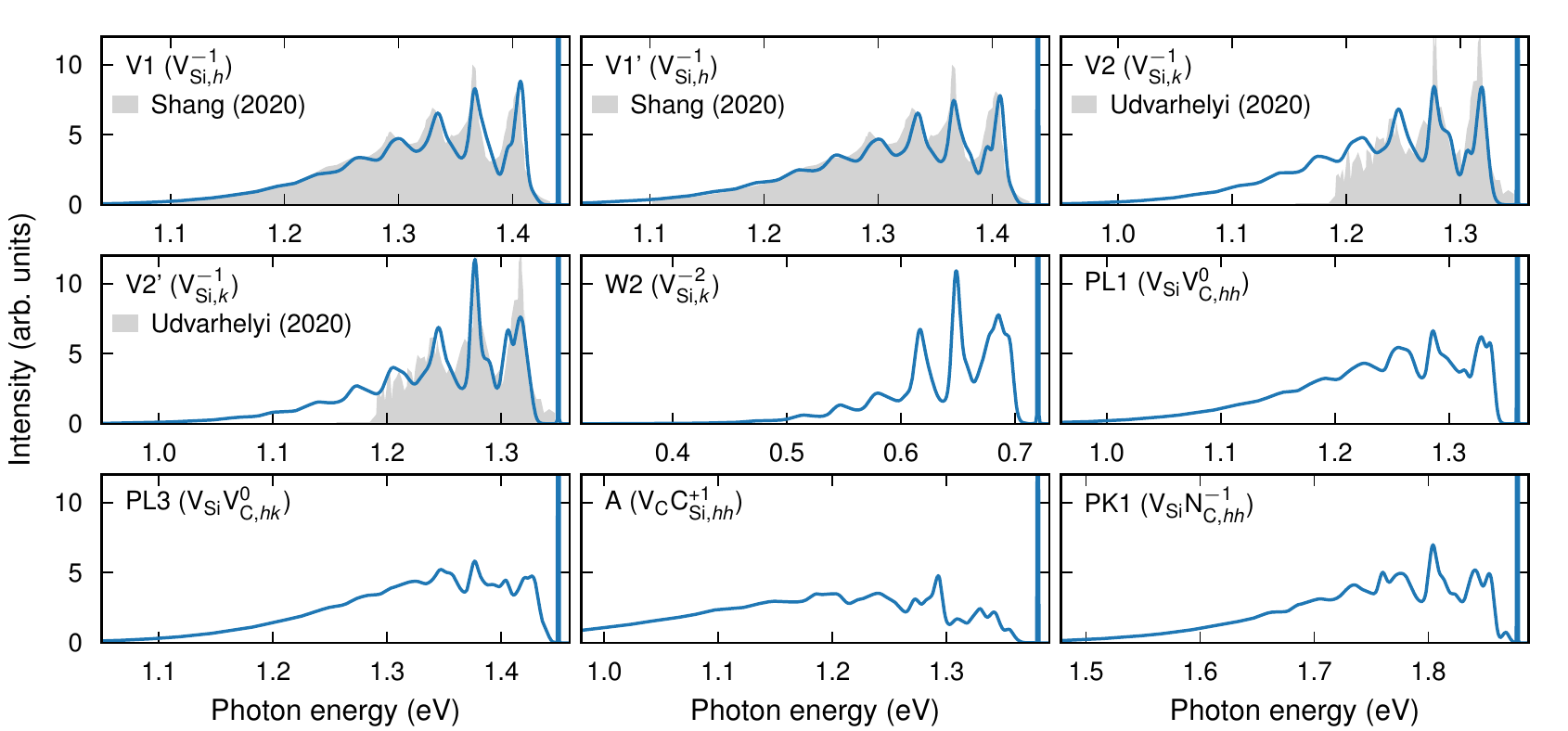}
    \caption{
        Emission lineshapes for the defect transitions considered in this study.
        The V1, V1' and V2, V2' (\vsineg) transitions are shifted to the respective experimental \glspl{zpl} in order to facilitate the comparison between calculation and experiment.
        Experimental data from Refs.~\citenum{Shang2020} (Shang, 2020) and \citenum{Udvarhelyi2020} (Udarvhelyi, 2020) are shown by the shaded areas.
        The intensity of the experimental lineshapes from Ref.~\citenum{Udvarhelyi2020} has been scaled to match the peaks of the computed lineshapes.
    }
    \label{fig:lineshapes-all}
\end{figure*}

The calculated emission lineshapes are shown in \autoref{fig:lineshapes-all},
where the \glspl{zpl} have been shifted to the respective experimental value to simplify the comparison where applicable.
The emission spectra consist of a sharp \gls{zpl} followed by a \emph{structured} \gls{psb} at lower energy.
It is evident from the results for the individual defects that the \glspl{psb} share some features:
(i) the \glspl{zpl} are separated from the \glspl{psb} by an energy gap of about \unit[30]{meV},
(ii) the broadening of the \glspl{psb} ranges from 200 to \unit[500]{meV}, corresponding to the respective \glspl{hrf} (\autoref{tab:info}), and
(iii) smaller \glspl{hrf} (number of created phonons) are associated with more structured \glspl{psb}.

The strength of the electron-phonon coupling can be partly assessed by the $Q^2$ value (\autoref{tab:info}), which measures the magnitude of the geometric difference between initial and final state.
It varies from 0.4 to \unit[1.0]{\qunit} for the W2 (\vsidblneg) and PL3 (\vcvsi) transitions showing the smallest and largest displacements, respectively.
The \glspl{dwf} varies from 2\% to 23\% with larger values implying a larger fraction of emission intensity in the \gls{zpl}.
By this measure the \vsidblneg{} center, which has  not been experimentally observed though, should be the best \gls{spe} candidate.
The V2' (\vsineg) transition results in a smaller lattice distortion at a value of \unit[0.47]{\qunit} compared with \unit[$0.61-0.67$]{\qunit} (\autoref{tab:info}) for the other transitions on \vsineg{}.
The PL1 (\vcvsi), PL3 (\vcvsi), and PK (\ncvsi) transitions exhibit the same $Q^2$ but differ in \glspl{hrf} (PK1 = PL1 $<$ PL3).
As a result, one observes rather similar lineshapes, especially between PL1 and PK1, except for a hump at about \unit[10]{meV} below the \gls{zpl} in the PK1 transition, which is due to a localized phonon.
Although the inclusion of an adjacent \vc{} to the \vsi{} in the transitions PL1 and PL3 modifies the position of the \gls{zpl}, it does not induce a dramatic change in the emission lineshape.
The optical emission lineshape of the transition A (\vccsi) with a tail of about \unit[500]{meV} shows the widest \gls{psb}.
We assign it to the contributions of high energy phonons and stronger coupling between electron and phonons during emission.

Our calculations for the V1/V1' and V2/V2' transitions involving \vsi{} defects are in excellent agreement with experimental data \cite{Shang2020, Udvarhelyi2020}.
They yield similar lineshapes and the \gls{psb} width is about \unit[300]{meV}.

The lineshape of the A transition has been measured \cite{Castelletto2013} at \unit[120]{K} and was hypothesized to originate from the \vccsi{} defect.
The computed lineshape does not agree favorably with that measurement, which poses the question whether the measured transition is actually taking place on \vccsi{}.
The comparison with experimental data can also be made by using the \gls{dwf} ($e^\text{-\gls{hrf}}$) values in \autoref{tab:info} to benchmark our calculations.
Only the experimentally reported \gls{dwf} for the A transition (1\%) \cite{Castelletto2013} deviates from our calculation (2.29\%), which we attribute to the finite temperature effect present in the experiments.
Overall, the calculations are in excellent agreement with experiments.

\subsection{Vibrational modes analysis}

\begin{figure}[ht!]
    \centering
    \includegraphics{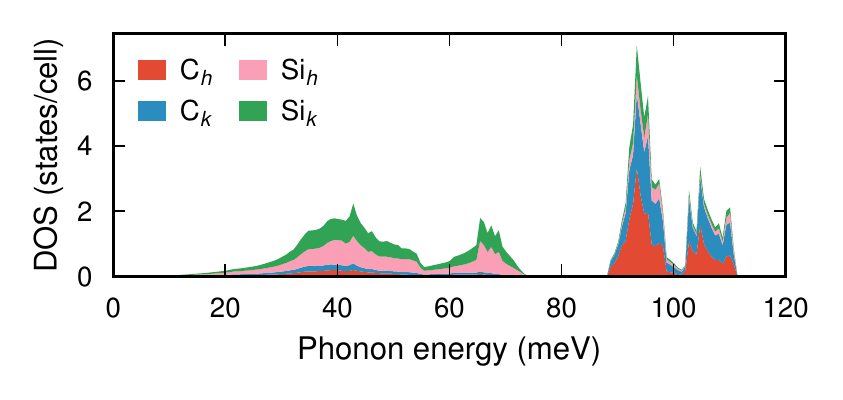}
    \caption{
        Partial phonon densities of states for the ideal structure of 4H-SiC.
    }
    \label{fig:dos}
\end{figure}
\begin{figure*}
    \centering
    \includegraphics{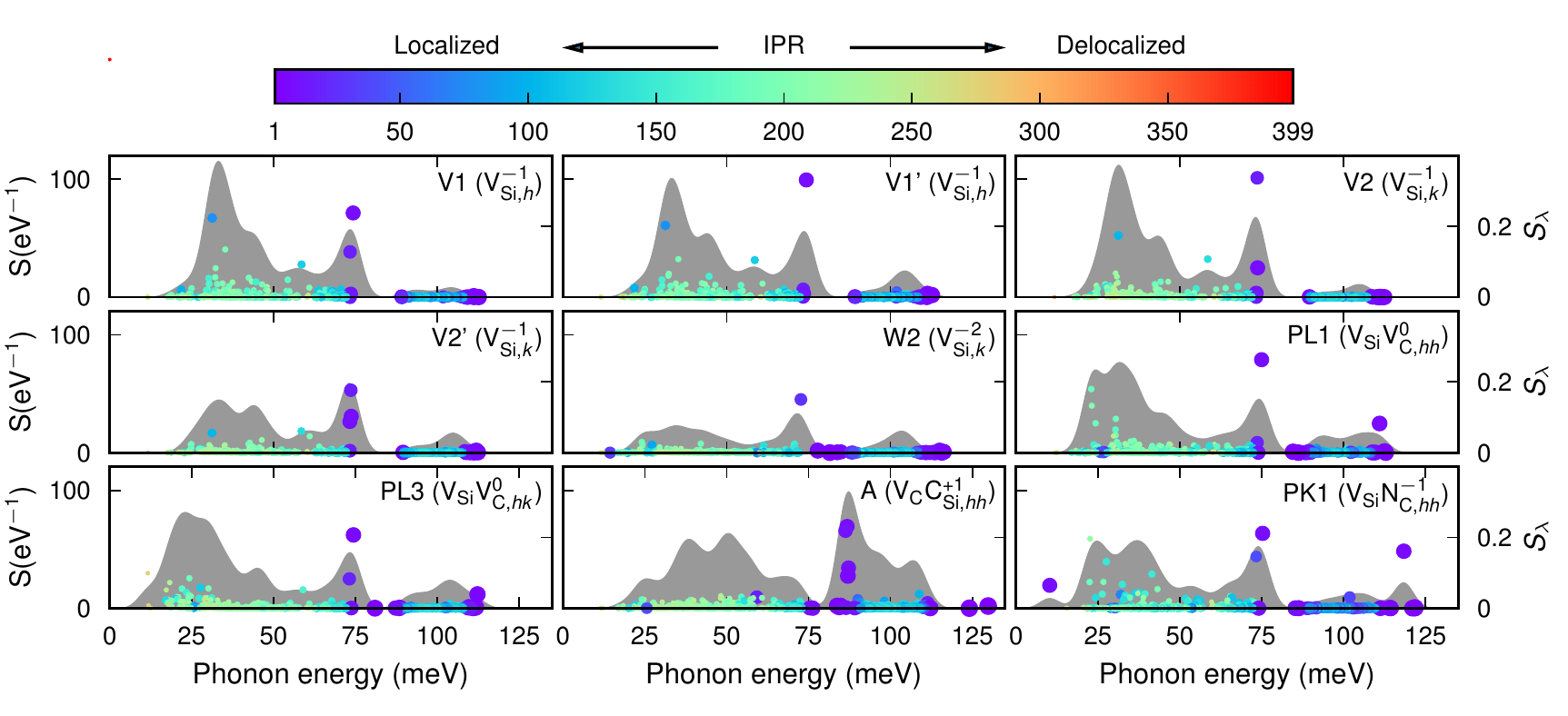}
    \caption{
        Spectral functions \specf{} (gray, shaded areas) and partial \glspl{hrf} (dots) for all considered defects.
        The \glspl{ipr} associated with each mode are indicated both by color and the size of the markers of the partial \gls{hrf}s.
    }
    \label{fig:hr-ipr}
\end{figure*}
\begin{figure*}[ht!]
    \centering
    \includegraphics[width=\linewidth]{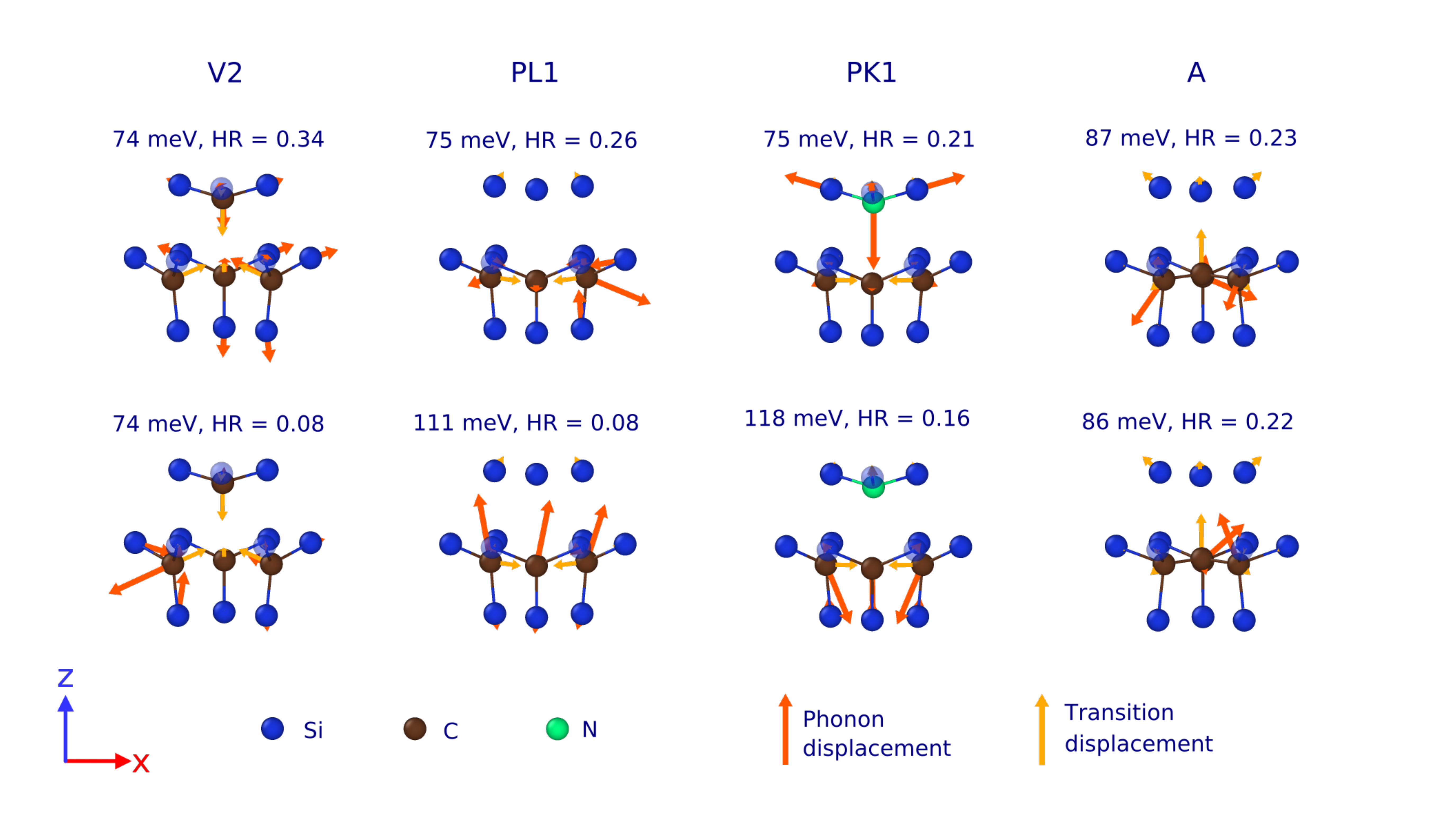}
    \caption{
        Transition displacement and phonon displacement vectors for a set of selected modes. The transition displacement vectors are amplified by a factor of 20 and the phonon displacement vectors are amplified by a factor of 5. The figure is created using the software {\scshape ovito} \cite{ovito}.
    }
    \label{fig:local-modes}
\end{figure*}
\begin{figure}[ht!]
    \centering
    \includegraphics[width=\linewidth]{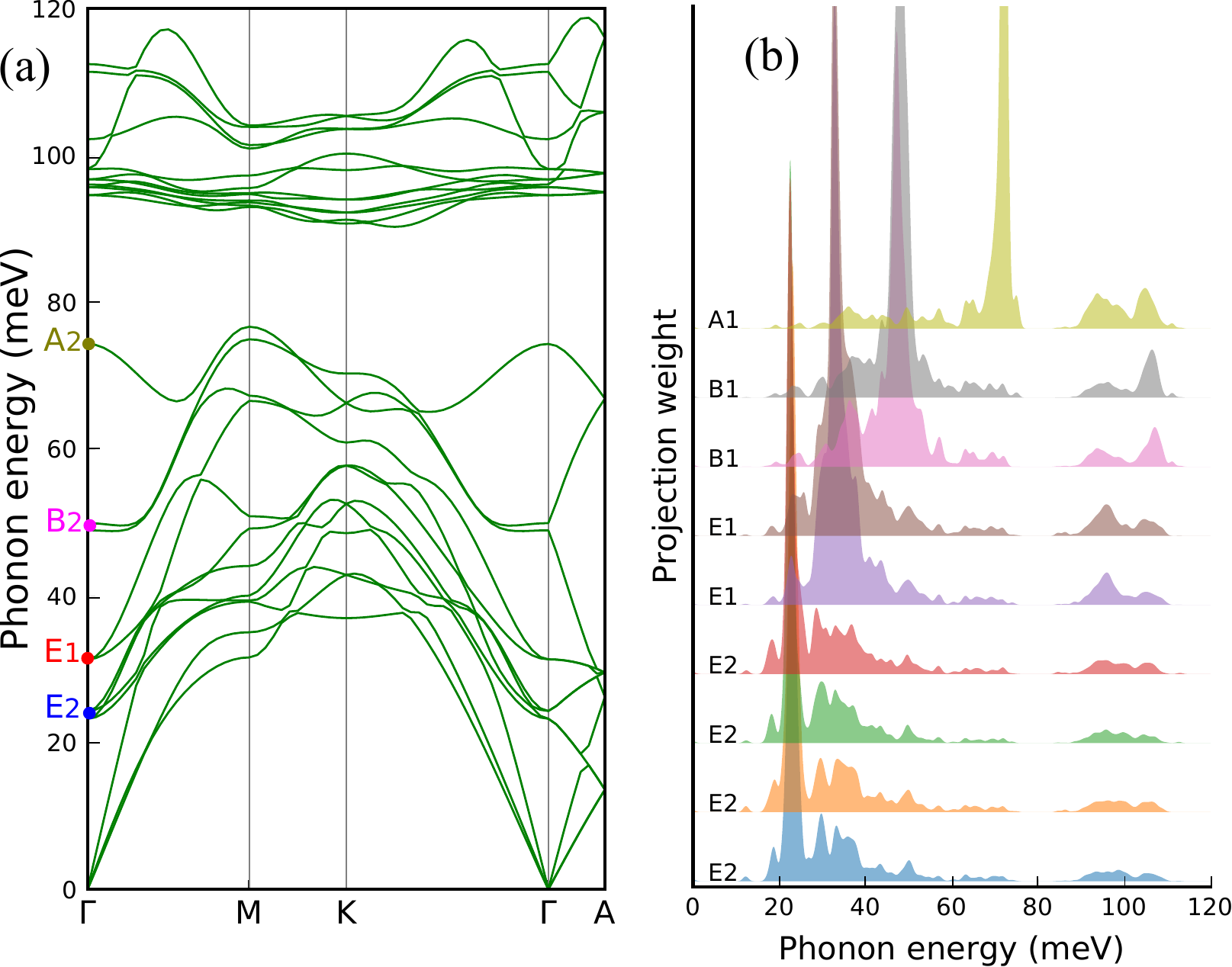}
    \caption{
        (a) Phonon dispersion of pristine 4H-SiC along high symmetry directions.
        (b) Projection of of (\vcvsi{})$^0_\text{hh}$ vibrational modes onto modes of the ideal structure (represented by its primitive unit cell).
        The configuration of defect is hh.
    }
    \label{fig:proj}
\end{figure}
Our phonon calculations for the pristine system using a \gls{puc} give the high energy Raman-active mode at \unit[96.63]{meV}, which is in excellent agreement with the experimental value of \unit[96.59]{meV} \cite{Wan2018}.
The phonon \gls{dos} can be divided into two parts (\autoref{fig:dos}).
In the energy range up to \unit[75]{meV}, the Si modes are dominant due to the larger mass of the Si atoms, while the double-peaked C-based modes are active at higher energies and centered at around \unit[100]{meV}.
The phonon band gap is about \unit[14.2]{meV} starting from \unit[76.4]{meV}.
Note that we see a slight difference between the partial \glspl{dos} of atoms placed at $h$ and $k$ sites due to the different arrangement of second nearest neighbors.

To gain insight into the contribution of different vibrational modes to the \glspl{psb}, we now consider the electron-phonon coupling spectral functions (\autoref{eq:s}).
To evaluate the localization of the modes in more details, we analyzed the \gls{ipr} for the mode $\lambda$ as \cite{Bell1970, Alkauskas2014}
\begin{equation}
    \mathrm{IPR}_{\lambda} = \frac{1}{\sum_a({\bf u}_{a,\lambda},{\bf u}_{a, \lambda})^2}.
\end{equation}
IPR$_{\lambda}=$ 1 if only one atom vibrates, while IPR$_{\lambda} = N$ if all $N$ atoms in the supercell vibrate with the same amplitude.
Moreover, the localized vibrational modes that give rise to strong \glspl{hrf} are collected and schematically illustrated.

The electron-phonon spectral functions are shown in \autoref{fig:hr-ipr}, accompanied by the plots of partial \glspl{hrf} and \gls{ipr}.
The spectral function \specf{} for the transitions on \vsineg{} (V1, V1', and V2) show only minor differences while the V2' transition can be clearly distinguished from the other three transitions.
This is reflected in the \gls{hrf}, which is about \unit[2.8]{} for V1, V1', and V2 while it is \unit[1.9]{} for V2'.

Overall, the vibrational contribution to the transitions can be divided into
(i) acoustic and optical bulk modes with energy less than \unit[45]{meV}, e.g., the V1, V1', and V2 transitions, and to a lesser extent the V2' transition, which exhibits a pronounced peak at \unit[35]{meV}.
This peak is composed of several modes with \gls{ipr}$>$100 (the maximal value of the \gls{ipr} for this cell is 399).
In other words, the peak is dominated by a single mode that has components of the bulk phonon displacement vector parallel to the carbon atoms adjacent to the vacancy site.
(ii) Defect modes that are mostly located around the lower edge of the phonon band gap.
In the V2 transition, the \unit[73]{meV} peak can be attributed to two modes with relatively low \gls{ipr} values of 8 and 18.
In V2', there are 3 modes (\gls{ipr} between 9 and 18) that have \glspl{hrf} between 0.09 and 0.18 making the total coupling in the region around \unit[73]{meV} similar in strength to the other transitions.
We note that there are many well localized modes (\gls{ipr}$\sim$ $4-6$) at the high frequency edge $\sim$\unit[110]{meV} in all transitions on \vsineg{}.
These modes do not however couple strongly to the transition displacement since the phonon displacement vectors are nearly perpendicular to the transition displacement.
The W2 transition (\vsidblneg) resembles the V2' spectral function with a weaker coupling to the vibrational degrees of freedom as indicated by the small total \gls{hrf} of 1.47.

The coupling to the vibrational modes is similar between the transitions on divacancies (PL1 and PL3) as evident from the spectral functions and the similar \glspl{hrf} of 2.75 and 3.08, respectively.
The general features of the spectral functions are similar to the transitions on \vsineg{},
except the broader peaks at 20 to \unit[50]{meV}.
The strongest coupling is at \unit[75]{meV} in both PL1 and PL3.
The mode is a low \gls{ipr} mode with values of 8 (7) in PL1 (PL3).
In addition, the spectral weight in the high frequency range, around \unit[100]{meV}, is slightly larger compared to the case of \vsineg{}. 
The peaks centered at \unit[23]{meV}, \unit[35]{meV}, and \unit[48]{meV} originate predominantly from several delocalized (bulk) modes with \gls{ipr}$>$100.
The coupling to the vibrational modes is, however, much larger for the A transition than other considered transitions.
The strongest coupling is to a set of four quasi-degenerate modes at the upper edge of the phonon band gap (\unit[$86-87$]{meV}) with partial \glspl{hrf} between 0.09 and 0.23.

For PK1 (\ncvsi), the coupling to the vibrational modes consists of three distinct bands at around \unit[$20-50$]{meV}, \unit[75]{meV} and \unit[120]{meV}.
The localized vibrational mode around \unit[10.3]{meV} is absent in other intrinsic centers.
Its \gls{ipr} value is 9.
The peak centered at \unit[75]{meV} results mainly from the coupling of two modes at 73 and \unit[75]{meV} with \glspl{hrf} of 41 and 8, respectively.

The selective vibrational modes and atomic displacements due to electron excitation are shown in \autoref{fig:local-modes}.
The dominant geometrical difference between the ground and excited states for all transitions on \vsineg{} is the radial motion (relative to the vacancy) of nearest neighbor carbon atoms, which displace by \unit[0.08]{\AA} in the case of V1 and \unit[0.05]{\AA} in the case of V1'.
The displacements of atoms after relaxation for these defects are inward upon emission as shown for V2.
These modes are quasi-degenerate and predominantly consists of motion of the nearest (relative to the vacancy) carbon atoms.
The intensity of the \unit[73]{meV} peak in the spectral function is, however, not weaker in the case of V2' compared with the other transitions (V1, V1', V2),
although the displacement of nearest neighbor carbon atom is smaller.
The phonon displacement for the \unit[75]{meV} localized mode is shown for the case of PL1.
During the transition, the closest C atoms to the silicon vacancy undergo the largest displacements towards the vacancy, while the Si atoms next to the C vacancy slightly move.
In addition, the atomic vibrations mainly happen in the Si vacancy plane. 
At around \unit[111]{meV}, there is a relatively weak, albeit stronger than in the case of \vsineg{}, coupling to modes with \glspl{ipr} of around 5.
The C atom displacements induced by transitions are almost perpendicular to vibrational motion.
The local modes in the A transition are distributed over relatively few atoms.
The C$_\mathrm{Si}$ atom displaces the most during the transition towards the vacant carbon atom upon emission.
Carbon atoms around and at the antisite are engaged in the atomic motions.
In the PK1 transition at \unit[75]{meV} the vibrational mode displaces the nearest Si atoms toward N.
The mode at \unit[118]{meV}, with an \gls{ipr} of 7, originates from the strong vibrations of C atoms positioned in the same SiC plane as the vacancy.
The structure relaxes around the vacant site.

To elucidate the symmetry of the participating phonons,
we can decompose the modes that predominantly contribute to the spectral function of PL1 (\autoref{fig:proj}).
To this end, we project the supercell (SC) modes, $\vb^{\rm SC}$, onto each of the PUC eigenmodes $\vb^{\rm PUC}$, at the $\Gamma$-point
\begin{equation}
   n (\omega) = \sum_{\lambda} {|\inner{\vb^{\rm PUC}}{{\vb^{\rm SC},\lambda}}|^2}~\delta(\omega - \omega_\lambda).
\end{equation}

This analysis allows us to identify the respective bulk phonons by inspecting the similarity between atomic vibrations contributing to the \gls{psb} and unfolded $\Gamma$-modes.
4H-SiC has point symmetry group $C_{6v}$ where A$_1$, B$_1$, E$_1$, and E$_2$ are the vibrational modes at the $\Gamma$-point \cite{Feldman1968}.
Both A$_1$ and E$_1$ modes are Raman- and IR-active, E$_2$ is only Raman-active, while B$_1$ is optically forbidden.
We assign the \unit[23]{meV} centered peak to the planar optical E$_2$ modes.
The doublets have a splitting of around \unit[1]{meV}.
The peak at \unit[35]{meV} can originate from a mixed symmetry of planar acoustic (at M) and optical modes with symmetry of E$_1$ \cite{Shang2020}.
The combination of axial optical and acoustic modes shape the spectral function peak at \unit[48]{meV}.
Indeed, we expect similar behaviour for other color centers.

\section{Conclusions}
In this work we have investigated the vibrational and associated optical properties of \vsineg{}, \vsidblneg{}, \vcvsi{}, \vccsi{}, and \ncvsi{} defects in 4H-SiC.
We compared the vibronic structure of different color centers
by evaluating several important characteristics such as \glspl{zpl}, \glspl{hrf}, \glspl{dwf} as well as the structure of the \glspl{psb}.
Our calculated \glspl{dwf} and \glspl{psb} are in excellent agreement with the available experimental data.
We found that the effect of charge state can be great on \gls{zpl} and \gls{dwf} of \vsi{} as it modifies the emission lineshape, whereas the lineshape does not differ between hh and hk pairs in \vcvsi{}. 
We identified few narrow peaks that correspond to multiphonon interactions.
The \gls{psb} is feature-rich for smaller \gls{hrf} values but shows a belly in the A transition due to strong electron-phonon coupling.
Further, we predicted few localized vibrational modes around the phonon band gap that notably contribute to the PL spectra.
Indeed, the electron transfer process to the ground state creates bulk acoustic phonons, though it is less-pronounced for \vccsi{} color center.
Overall our work thereby provides a reference database for single photon sources and spin qubits in 4H-SiC.

\section{Acknowledgements}
This work has been supported by the Academy of Finland under Project No.311058 and the Knut and Alice Wallenberg Foundation (2014.0226).
T. A.-N. has been supported in part by the academy of Finland QTF CoE grant No. 312298.
We also thank CSC-IT Center Science Ltd. (Finland) and the Swedish National Infrastructure for Computing at PDC (Stockholm, Sweden) for generous grants of computer time.
The authors also would like to thank professor Martti Puska for his support.
\bibliography{ars}
\end{document}